\documentclass[a4paper,12pt]{article}

\usepackage[margin=2.5cm]{geometry}

\usepackage{textgreek}

\usepackage[all]{nowidow}
\usepackage{amsmath}
\usepackage[version=4]{mhchem}
\usepackage{siunitx}
\usepackage{tabularx}
\usepackage{booktabs}
\usepackage{float}
\usepackage{graphicx}
\usepackage{caption}
\usepackage{subcaption}
\usepackage[
backend=biber,
style=vancouver
]{biblatex}

\renewbibmacro*{volume+number+eid}{%
    \printfield{volume}%
    \printfield{number}%
}\DeclareFieldFormat[article]{number}{\mkbibparens{#1}}

\DeclareFieldFormat*{pages}{\addspace\mkcomprange{#1}}

\DeclareFieldFormat[article,periodical]{volume}{\mkbibbold{#1}}

\DeclareSIUnit \nucleon {u}
\DeclareSIUnit \ion {ion}
\DeclareSIUnit \frame {frame}
\DeclareSIUnit \Hit {Hit}
\DeclareSIUnit \vertices {vertices}
\DeclareSIUnit \vertex {vertex}

\usepackage{hyperref} % Load last

\graphicspath{{.}}

\newcolumntype{C}{>{\centering\arraybackslash}X}

\newcommand{\cbase}{% Prevent leading space
    \ce{^12C}% Prevent trailing space
}
\newcommand{\cion}{% Prevent leading space
    \ce{^12C^{6+}}% Prevent trailing space
}

\newcommand{\singleFig}[5][!htbp]{%
    \begin{figure}[#1]%
        \centering%
        \captionsetup{width=15cm}%
        \includegraphics[width=15cm]{#2}%
        \caption[#5]{#4}%
        \label{fig:#3}%
    \end{figure}%
}

\newcommand{\doubleFig}[6][!htbp]{%
    \begin{figure}[#1]%
        \centering%
        \captionsetup{width=16cm}%
        \begin{subfigure}[t]{7.9cm}%
            \centering%
            \includegraphics[width=\linewidth]{#2}%
            \phantomsubcaption%
            \label{fig:#4_a}%
        \end{subfigure}%
        \hspace{0.2cm}%
        \begin{subfigure}[t]{7.9cm}%
            \centering%
            \includegraphics[width=\linewidth]{#3}%
            \phantomsubcaption%
            \label{fig:#4_b}%
        \end{subfigure}%
        \caption[#6]{#5}%
        \label{fig:#4}%
    \end{figure}%
}

\newcommand{\figref}[1]{% Prevent leading space
    Figure \ref{fig:#1}% Prevent trailing space
}

\newcommand{\tabref}[1]{% Prevent leading space
    Table \ref{tab:#1}% Prevent trailing space
}

\begin{document}
    \begin{center}
        \Large{Clinical beam test of inter- and intra-fraction relative range monitoring in carbon ion radiotherapy}
    \end{center}
    \begin{center}
        \textbf{Devin Hymers}\textsuperscript{1},
        \textbf{Sebastian Schroeder}\textsuperscript{1},
        \textbf{Olga Bertini}\textsuperscript{2},
        \textbf{Stephan Brons}\textsuperscript{3},
        \textbf{Johann Heuser}\textsuperscript{2},
        \textbf{Joerg Lehnert}\textsuperscript{2},
        \textbf{Christian Joachim Schmidt}\textsuperscript{2},
        \textbf{and}
        \textbf{Dennis M\"ucher}\textsuperscript{1}
    \end{center}
    
    \textsuperscript{1}Institute for Nuclear Physics, University of Cologne, 50937 Cologne, Germany
    
    \textsuperscript{2}GSI GmbH, 64291 Darmstadt, Germany
    
    \textsuperscript{3}{Heidelberg Ion Beam Therapy Centre (HIT), 69120 Heidelberg, Germany}
    
    \section{Abstract}
    \label{sec:ivi_abstract}
    
    Interaction Vertex Imaging (IVI) is used for range monitoring (RM) in carbon ion radiotherapy. The purpose of RM is to measure the Bragg peak (BP) position for each contributing beam, and detect any changes. Currently, there is no consensus on a clinical RM method, the use of which would improve the safety and consistency of treatment. The prototype filtered IVI (fIVI) Range Monitoring System is the first system to apply large-area and high-rate-capable silicon detectors to IVI. Two layers of these detectors track prompt secondary fragments for use in RM. This device monitored \qty{16}{\cm} and \qty{32}{\cm} diameter cylindrical plastic phantoms irradiated by clinical carbon ion beams at the Heidelberg Ion Beam Therapy Center. Approximately 20 different BP depths were delivered to each phantom, with a minimum depth difference of \qty{0.8}{\mm} and a maximum depth difference of \qty{51.9}{\mm} and \qty{82.5}{\mm} respectively. For large BP range differences, the relationship between the true depth difference and that measured by fIVI is quadratic, although for small differences, the deviation from a linear relationship with a slope of 1 is negligible. RM performance is strongly dependent on the number of tracked particles, particularly in the clinically-relevant regime. Significant performance differences exist between the two phantoms, with millimetric precision at clinical doses being achieved only for the \qty{16}{\cm} phantom. The performance achieved by the prototype fIVI Range Monitoring System is consistent with previous investigations of IVI, despite measuring at more challenging shallow BP positions. Further significant improvements are possible through increasing the sensitive area of the tracking system beyond the prototype, which will both allow an improvement in precision for the most intense points of a scanned treatment plan and expand the number of points for which millimetric precision may be achieved.
    
    \section{Introduction}
    \label{sec:ivi_intro}
    
    Ion-beam radiotherapy is a growing alternative to the conventional photon beam for external-beam radiotherapy \autocite{malouff_carbon_2020}. Key advantages of irradiation with ion beams include the precise dose delivery, facilitating treatment using a highly conformal dose distribution \autocite{tessonnier_experimental_2017}, and the higher biological effectiveness \autocite{tinganelli_carbon_2020}. While the most common form of ion-beam therapy is the proton, the benefits of ion-beam therapy scale with the ion mass, leading to the development and implementation of carbon-ion radiotherapy (CIRT). Although CIRT is more expensive than other forms of external-beam radiotherapy, its sharper dose gradients may facilitate better sparing of healthy tissue, including radiosensitive organs at risk, which is of particular benefit in treatment of pediatric patients with the greatest potential for a long post-treatment lifespan \autocite{laprie_paediatric_2015, peeters_how_2010}. Heavier ions have also shown some enhancement in the ability to treat hypoxic or otherwise radioresistant tumours \autocite{sokol_carbon_2023}.
    
    A defining characteristic of ion-beam radiotherapy (including CIRT) over other types of external beam radiotherapy is its inverted dose distribution, related to the variable stopping power of the \cion{} beam as it loses energy quasi-continuously in the patient. This dose distribution produces a low entrance dose, rising to a sharp maximum at the beam endpoint, or Bragg peak (BP), followed by a very low exit dose attributable to lighter particles produced by rare fragmentation of the primary \cbase{} atoms \autocite{amaldi_radiotherapy_2005}. The high mass of the \cbase{} nucleus also reduces the lateral spread of the beam as compared to lighter ions, yielding overall a well-defined beamspot with relatively sharp edges, even at significant depth within the patient \autocite{tessonnier_experimental_2017}. As the typical target volume containing a tumour is much larger than a single beamspot, a typical irradiation consists of many independent beams, delivered sequentially to different positions and with a variety of energies, resulting in uniform delivery of the prescribed dose to the entire target volume \autocite{kramer_treatment_2000, haberer_magnetic_1993}. The full prescribed radiation dose is commonly split into fractions, with one fraction delivered each day or each weekday, for several weeks. 
    
    However, the benefits brought by the BP also pose an additional challenge for ion-beam radiotherapy. Because the dose distribution is so sensitive to the beam range, and therefore the beam energy, there is a demand for range monitoring (RM) methods, which quantify the beam range. Currently, uncertainties in beam delivery demand a safety margin around the tumour, to account for possible errors in BP depth \autocite{andreo_clinical_2009}. A typical value for a safety margin in the adult head is around \qty{3}{\mm}, which results in a significant volume of healthy tissue receiving the full prescribed dose, in addition to the tumour itself \autocite{kelleter_-vivo_2024}. If the beam position could be monitored with sufficient precision, this margin could be reduced, sparing healthy tissue and reducing the risk of secondary tumours, while maintaining the same rate of tumour control.
    
    RM may provide either an absolute measure, which may be directly mapped to a BP position or dose distribution on an anatomical image such as a planning CT, or a relative measure, which yields only the relative shift in BP position or dose distribution between two irradiations. For relative measures, the comparison may be made between two irradiations which are part of the same fraction (i.e. intra-fraction), or between two irradiations which are part of different fractions (i.e. inter-fraction). Although relative RM alone cannot determine whether a treatment has properly covered the tumour region, it is capable of detecting and measuring deviations in patient positioning or internal structure which occur during irradiation (intra-fraction) or between irradiations (inter-fraction). Inter-fraction changes, which can occur due to variations in patient setup and alignment, or changes to internal structure such as different positioning of organs or filling of cavities, can lead to systematic positioning errors which invalidate a treatment plan, and indicate the need for an updated treatment plan \autocite{fischetti_inter-fractional_2020}. Intra-fraction changes, which may be due to periodic processes such as respiration, or aperiodic movement such as a muscle spasm, do not necessarily invalidate the entire treatment plan, but require accommodation or correction to ensure that the entire target volume is treated uniformly \autocite{ammazzalorso_dosimetric_2014}.
    
    Although a number of RM methods have been studied, no method has yet undergone widespread integration into clinical practice. Many of these methods rely on secondary radiation produced by interactions between patient tissue and the therapy beam, to produce some signal which can be measured by external detectors. The most mature of these techniques is based around positron emission tomography (PET), which depends on the activation of patient tissue or beam-like fragments which decay by positron emission to produce a signal which is correlated to beam position \autocite{parodi_vivo_2018}. Initial PET RM investigations focused on post-irradiation measurement, as the signal is delayed by the lifetime of the activated fragments. However, the time required to transport the patient to a PET scanner proved to limit the collectible data, and could only provide an integrated image of the entire fraction, rather than independent measurement of each BP position. Further work has brought PET systems into the treatment room, including integration into the treatment table (in-beam PET), which provides a faster and more complete image, as well as providing some intra-fraction monitoring \autocite{handrack_sensitivity_2017, pennazio_carbon_2018}. In-beam PET for CIRT is currently undergoing clinical trials at the Italian National Center of Oncological Hadrontherapy (CNAO, Pavia, Italy), which has so far achieved a range difference precision of \qty{3.7}{\mm} for tumours of the head and neck \autocite{kraan_-beam_2024}.
    
    A second option for prompt monitoring, which requires a less bulky detector than PET, and can collect all data online during irradiation, is prompt gamma (PG) imaging. This method, which monitors secondary gamma ray photons, is not limited to the photons produced by positron emission, but is sensitive to photons from all processes, increasing the number of useful interactions \autocite{parodi_vivo_2018}. However, the challenge with PG methods is to correlate a photon detection to the site of its production, as a single gamma ray interaction does not inherently carry information about the trajectory of the incoming photon. This information may be added by collimating the detector, constraining the origin of the photon to a narrow slice of the patient directly in front of the detection system. However, the thick collimators required for effective attenuation significantly reduce the detection efficiency of PG systems \autocite{testa_dose_2009}. There have been recent successes in PG RM for proton radiotherapy; however, it is challenging to apply the same methods to CIRT, as the gamma ray production cross sections are similar, but the number of primary ions needed to deliver the same dose is orders of magnitude lower in CIRT than in proton therapy. Therefore, an efficiency increase by a factor of \num{20} would be needed to achieve the same precision in CIRT \autocite{krimmer_prompt-gamma_2018}.
    
    Another RM method which uses only prompt radiation is interaction vertex imaging (IVI), which is the focus of this work. Rather than monitoring prompt photons as in PG, IVI uses charged particles, typically protons \autocite{amaldi_advanced_2010}. Although the fraction of secondary protons which are able to exit the patient is much smaller than the fraction of secondary photons, the detection efficiency of charged particles is much greater, approaching \qty{100}{\percent} even for thin sensors. It is also straightforward to perform multiple detections with charged particles, using two or more detector layers in a tracker, allowing determination of direction of travel without requiring efficiency-reducing collimators. This information, along with the orientation of the treatment beam, is used to reconstruct the site of the fragmentation reaction (the interaction vertex) which produced the secondary particle \autocite{henriquet_interaction_2012}. Although the tracking and reconstruction process is more straightforward than with PG, the resolution of IVI is limited by the multiple Coulomb scattering of fragments as they exit the patient, reducing the precision with which a single secondary fragment can be reconstructed \autocite{ghesquiere-dierickx_investigation_2021}. A further limitation is the continual energy loss of the secondary fragments as they exit the patient: although the energy loss per unit distance is less than the primary treatment beam, the requirement that these secondary fragments must not stop before exiting the patient means that IVI is not sensitive to direct beam-patient interactions occurring at or immediately superficial to the BP \autocite{henriquet_interaction_2012}. A number of simulation and phantom-based studies have examined the IVI technique, with the most recent work focusing on relative RM and the detection of BP range differences \autocite{ finck_study_2017, gwosch_non-invasive_2013, felix-bautista_experimental_2019, hymers_intra-_2021}.
    
    A number of IVI devices are currently engaged in the clinical trial phase. A monolithic \qtyproduct[product-units=power]{30x30x30}{\cm} scintillator-based `Dose Profiler' was developed at CNAO, and is currently being tested in conjunction with the in-beam PET system \autocite{fischetti_inter-fractional_2020}. Another device, based on seven small silicon trackers, each with a sensitive area of \qtyproduct[product-units=power]{2.8x1.4}{\cm} is being studied at the Heidelberg Ion-Beam Therapy Center (HIT, Heidelberg, Germany) \autocite{kelleter_-vivo_2024}. Both of these systems are being studied for relative RM in tumours of the head and neck.
    
    An intermediate approach, previously proposed by the authors, uses a larger monolithic tracker, similarly to the device tested at CNAO, but comprised of thin silicon detectors, as with the device tested at HIT. In addition, the implementation of IVI used with this approach employs a software filter, which rejects secondary particles for which multiple Coulomb scattering has a significant impact on the accuracy of interaction vertex reconstruction. This implementation, called filtered IVI (fIVI), has shown promise as a high-precision relative RM method in previous subclinical tests \autocite{hymers_intra-_2021}.
    
    This work describes the first tests using clinical beams of the prototype fIVI Range Monitoring System, designed to support the use of high-precision fIVI under clinical conditions, and with the aim to achieve millimetric precision in relative RM. Achieving a large and highly rate-capable sensitive area with a single monolithic tracker allows great flexibility in positioning the device around a patient, to account for variations in size or tumour position, as opposed to an array of smaller trackers, which is more restrictive due to the common focal point. The large sensitive area also maximizes the data collected, which is important to perform RM subject to the dose constraint of clinical irradiation. A high-rate digital readout system is also required, to effectively distinguish events and perform tracking in the challenging environment of the treatment room. These tests, in homogenous phantoms with comparable dimensions to a human head and a human torso, represent the first application of large-area silicon sensors to IVI, as well as the first direct comparisons between clinical beams with BP depths significantly less than the phantom radius for phantoms differing in size by a factor of two.
    
    \section{Materials and Methods}
    \label{sec:ivi_materials}
    
    \subsection{Prototype fIVI Tracker}
    \label{sec:ivi_tracker}
    
    The functional unit of the prototype fIVI Range Monitoring System is the sensor module, with each module containing a single silicon sensor and front-end readout electronics. These modules were designed and manufactured at the University of Cologne, to accept the large-area silicon sensors and coupled readout electronics developed for the Silicon Tracking System of the Compressed Baryonic Matter experiment, which will be installed at the Facility for Antiproton and Ion Research (FAIR, GSI, Darmstadt, Germany) \autocite{heuser_technical_2013}. To maximize coverage of the detection system, an important feature for IVI RM, the largest sensor sizes were selected for this setup, with sensitive areas \qtyproduct[product-units=power]{6.0x6.0}{\cm} and \qtyproduct[product-units=power]{6.0x12.0}{\cm}.
    
    Each of these sensors is segmented to achieve position sensitivity, with \num{1024} segments per planar side. To facilitate detector arrays consisting of multiple adjacent sensors, all connections to each sensor are made along a single \qty{6.0}{\cm} edge; both sensor sides are segmented with a \qty{58.6}{\micro\m} pitch. Segments on opposite faces of the sensor are therefore non-orthogonal, with conventional axially-oriented strips on the n side of the sensor (orthogonal to the readout edge), and strips oriented at a \ang{7.5} angle from this axis on the p side of the sensor. Consequently, the spatial resolution of these sensors is orientation-dependent, with maximum resolution in the direction parallel to the readout edge.
    
    Sensor readout is accomplished using a fully digital, ASIC-based system, with each ASIC directly connected to the sensor by use of analog microcables. Each segment operates in an independent, self-triggered mode. Timing uses an \qty{80}{\mega\Hz} clock, with the clock incrementing on both the rising and falling edge. Therefore, the effective tick rate of the clock is \qty{160}{\mega\Hz}, allowing timestamps for trigger events on each segment to be assigned with \qty{6.25}{\nano\s} precision. Energy resolution is limited to 5 bits, in support of achieving fast timing performance.
    
    Two modules are combined to form the prototype tracker setup, with a  \qtyproduct[product-units=power]{6.0x6.0}{\cm} sensor placed \qty{12.0}{\cm} in front of a \qtyproduct[product-units=power]{6.0x12.0}{\cm} sensor. The use of a larger sensitive area for the rear layer increases the acceptance of the tracker, improving the proportion of secondary protons which can be tracked relative to a setup with a smaller rear sensor, and therefore increasing the total number of particles which can be tracked during a CIRT treatment. To protect the sensors and readout electronics from visible light and other electromagnetic interference, a \qty{1.0}{\mm} thick aluminum enclosure, shown in \figref{tracker}, covers both modules, forming a Faraday shield. A thinner \qty{16}{\micro\m} aluminum panel forms an entrance window, \qty{3.35}{\cm} in front of the first sensor layer.
    
    \doubleFig[!t]{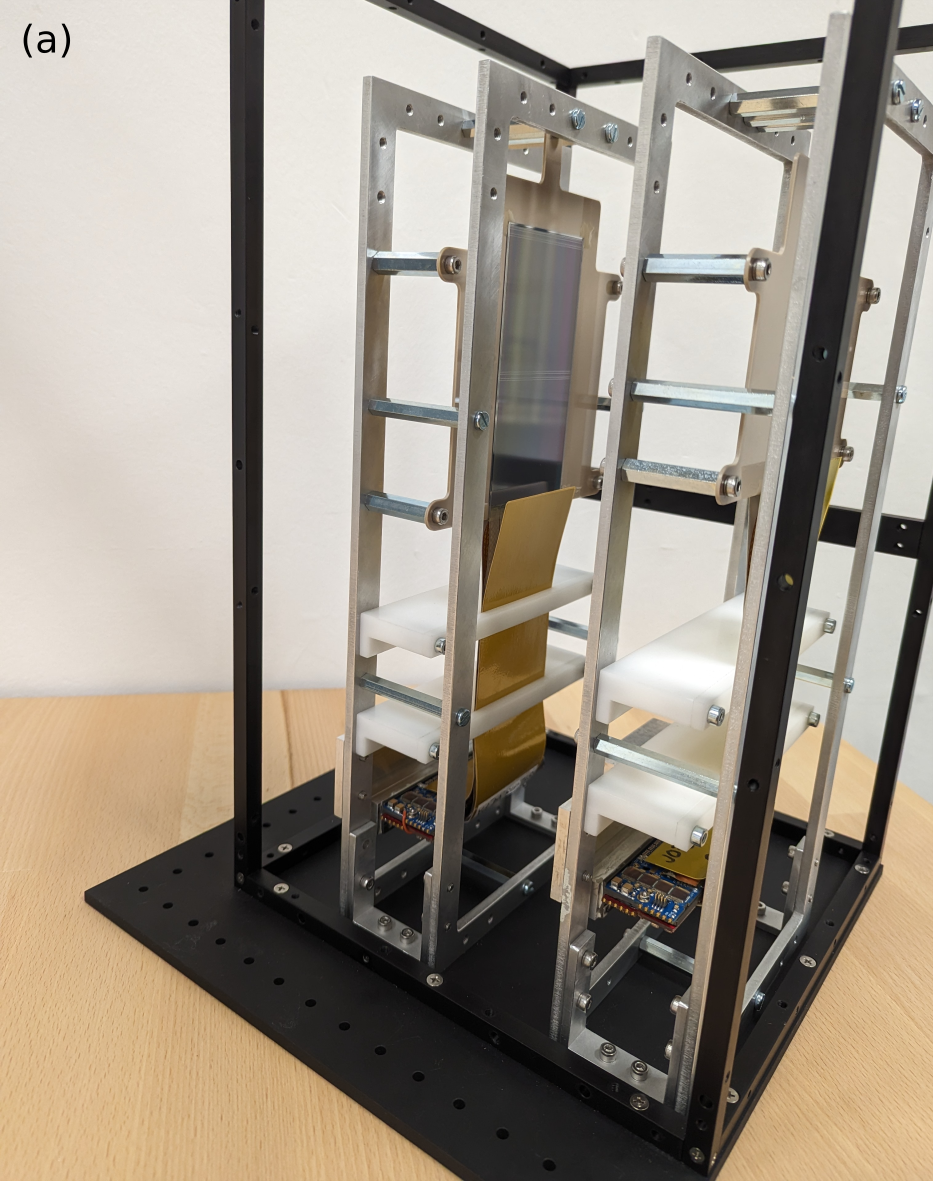}{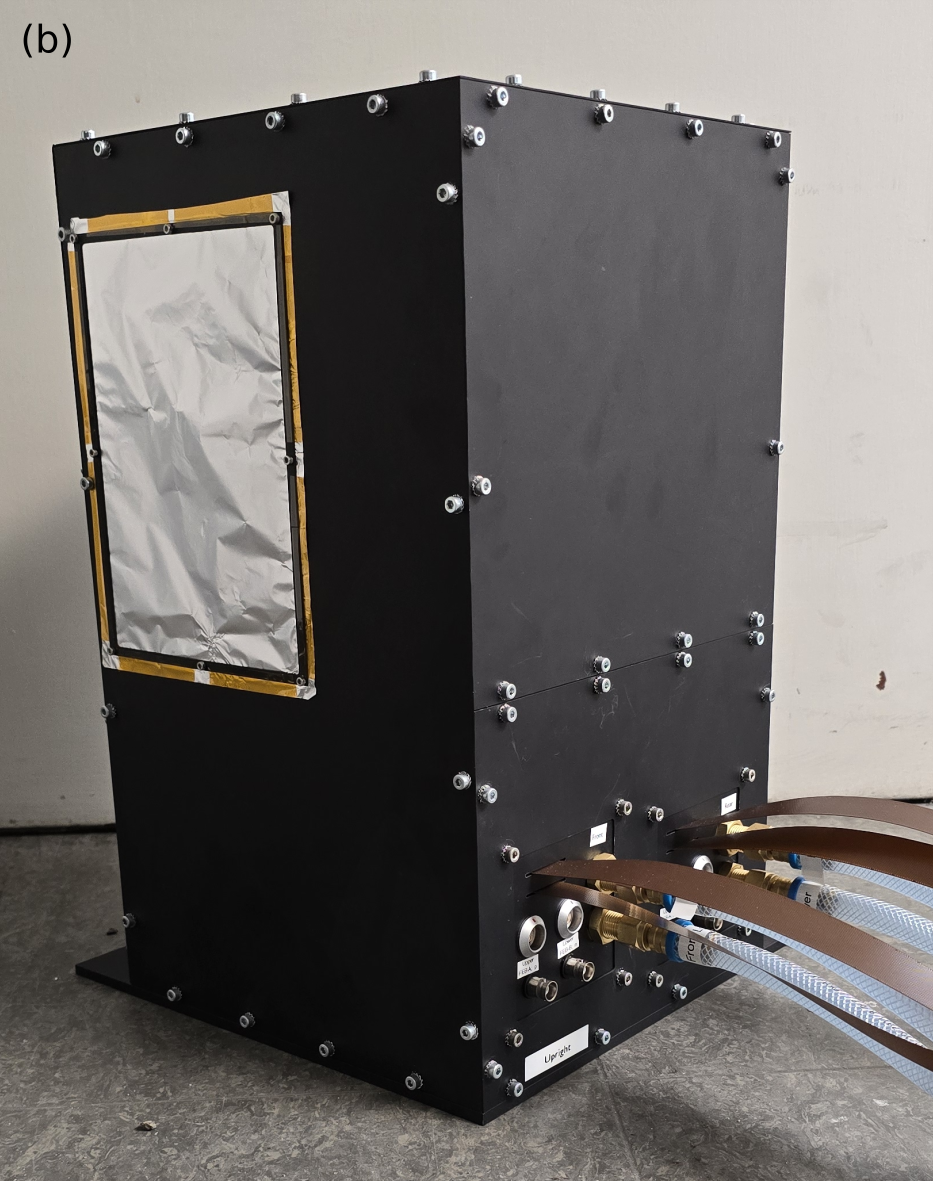}{tracker}{The assembled tracker setup. (a) Sensor modules mounted, without enclosure installed. Front-end readout electronics are visible at the bottom of each module. The baseplate region exterior to the frame provides a mounting surface to install the tracker in an experimental setup. (b) Enclosure installed. A foil entrance window (left) sits in front of the sensor modules. Connections for power, data, and liquid cooling of each module pass through the enclosure (lower right).}{The assembled tracker setup.}
    
    A more detailed description of the hardware and readout electronics may be found in a separate publication \autocite{hymers_evaluation_2025}.
    
    \subsection{Experimental Setup}
    \label{sec:ivi_setup}
    
    The prototype fIVI Range Monitoring System was installed at the quality assurance beamline of HIT. A cylindrical PMMA (poly methyl-methacrylate) phantom manufactured by PTW Freiburg GmbH (PTW T40027) was used as the primary irradiation target. This phantom had been previously measured to have a stopping power of \num{1.165} relative to water, corresponding to a density of approximately \qty{1.20}{\gram\per\cubic\cm}. This phantom consisted of two main parts: a \qty{16}{\cm} diameter core; and a \qty{32}{\cm} diameter outer shell with a \qty{16}{\cm} diameter central cutout, which fit snugly around the \qty{16}{\cm} core to produce a uniform \qty{32}{\cm} diameter cylinder. Each piece of the phantom contained four small cylindrical cutouts of \qty{9}{\mm} diameter, located near the outer edge of the phantom. These cutouts were filled with snugly-fitting PMMA inserts during irradiation. The \qty{16}{\cm} phantom contained one additional cutout at the center of the phantom; the insert in this position was used to fix the phantom in place during irradiation. Horizontal and vertical alignment markings on the phantom were used, along with the HIT optical alignment system, to colocate the center of the phantom with the beam delivery isocenter. This isocenter was used as the origin of the lab frame coordinate system throughout analysis.
    
    \doubleFig[t]{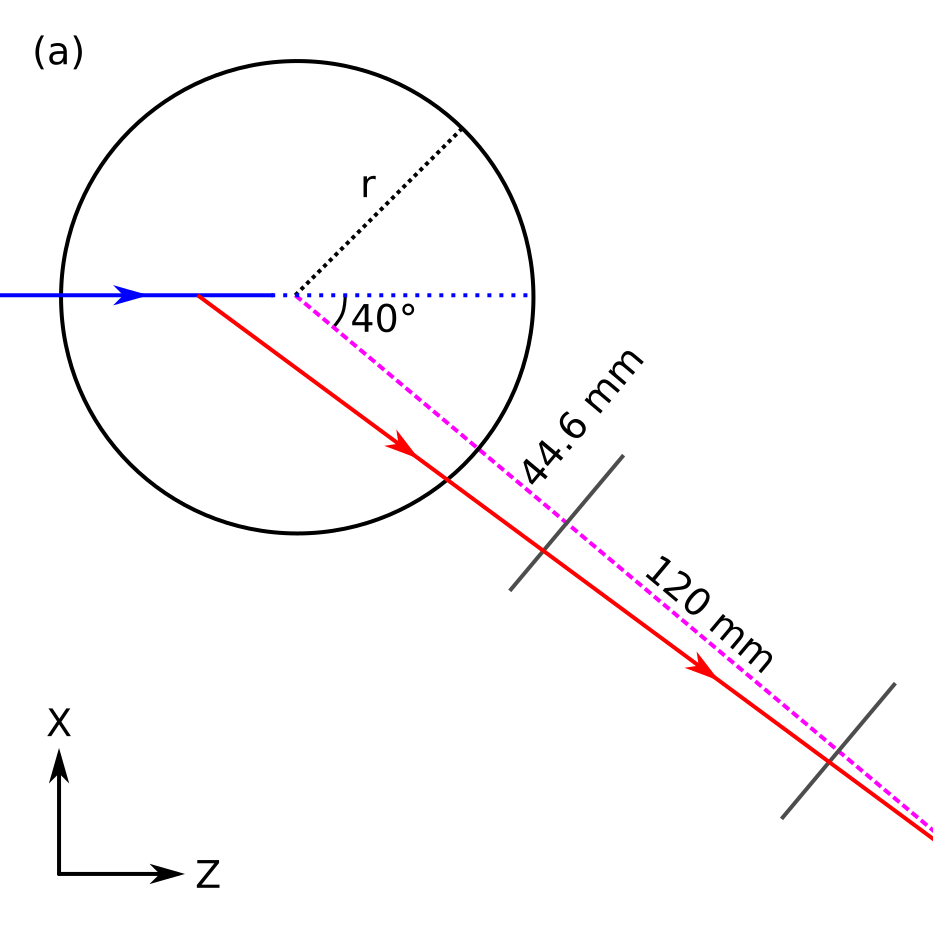}{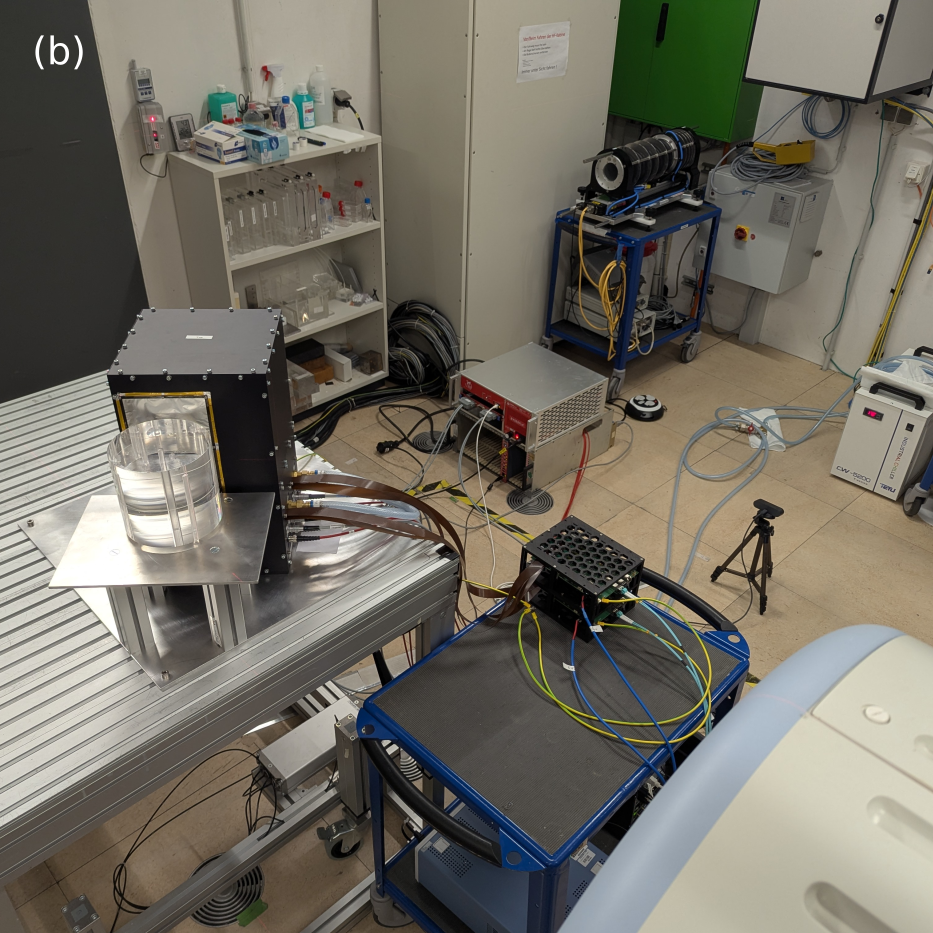}{setup}{Experimental setup for fIVI measurement. (a) Schematic diagram. The primary \cion{} beam (blue) was oriented along the $Z$ axis, travelling in the $+Z$ direction. The central axis of the tracker (magenta, dashed) was placed at a \ang{40} angle from the beam axis (blue, dotted), with the intersection of the two being at the center of the phantom (white). Sensors (grey) were placed orthogonal to the tracker axis, \qty{44.6}{\mm} and \qty{164.6}{\mm} from the edge of the phantom. Secondary fragments (red) were produced by beam-patient interactions; those fragments passing through both sensor layers were tracked. Data were collected for phantom radii $r$ of \qty{80}{\mm} and \qty{160}{\mm}. (b) Photograph of experimental setup. The beam exited the nozzle in the lower right, and travelled toward the setup (left), consisting of the phantom (clear plastic) and tracker (black). The smaller phantom, with a \qty{16}{\cm} diameter, is pictured.}{Experimental setup for fIVI measurement.}
    
    The tracker was placed at a \ang{40} angle from the beam axis, as shown in \figref{setup}, with its position fixed relative to the phantom by mounting both to the same aluminum plate. The central axis of the tracker was aligned to point at the center of the phantom. While previous studies have concluded that smaller off-axis angles may provide increased precision in reconstruction \autocite{finck_study_2017, ghesquiere-dierickx_investigation_2021}, these studies were performed for high depth, where the exit path length of forward-focused fragments differs more significantly from fragments at larger angles, and for small sensors, where the field of view at larger angles fails to cover the entire phantom. Therefore, new calculations were needed to support the current setup. For the current measurement, \ang{40} was selected as an angle which was expected to provide sufficient statistics for the previously proposed clinical fIVI tracker \autocite{hymers_intra-_2021}. Two positions were selected for the tracker, which placed the first sensor \qty{4.5}{\cm} from the edge of the \qty{16}{\cm} and \qty{32}{\cm} phantoms respectively. Again, although these positions were slightly closer than suggested by previous optimizations \autocite{finck_study_2017}, the wider field of view provided by the prototype fIVI tracker’s large sensors and the expected improvement from collection of a greater fraction of secondary protons supported this choice. In particular, for the \qty{16}{\cm} phantom, the selected distance and off-axis angle are similar to mini-trackers 1 and 2 of the monitoring system of \textcite{kelleter_-vivo_2024}. Due to the large sensitive area of both sensor layers, the field of view was still large enough to cover the entire relevant region of the phantom, with good acceptance, as seen in \figref{acceptance}. For the \qty{16}{\cm} phantom, the entire length of the beam axis was covered by the field of view, while for the \qty{32}{\cm} phantom, the field of view covered up to a depth of approximately \qty{31}{\cm}, i.e. from \qty{-16}{\cm} to \qty{+15}{\cm} along the $Z$ axis. However, since the largest BP depth tested was only approximately \qty{16}{cm}, close to the center of the phantom, the minor field of view limitation does not appreciably affect reconstruction.
    
    \doubleFig[t]{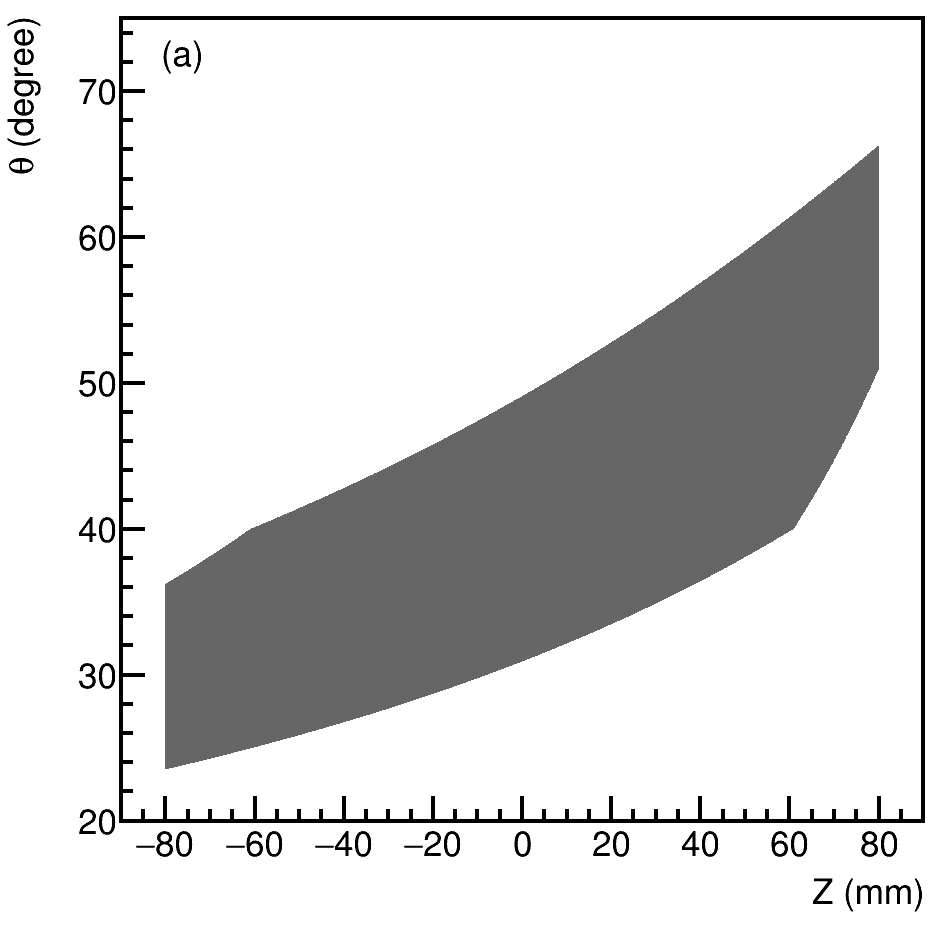}
    {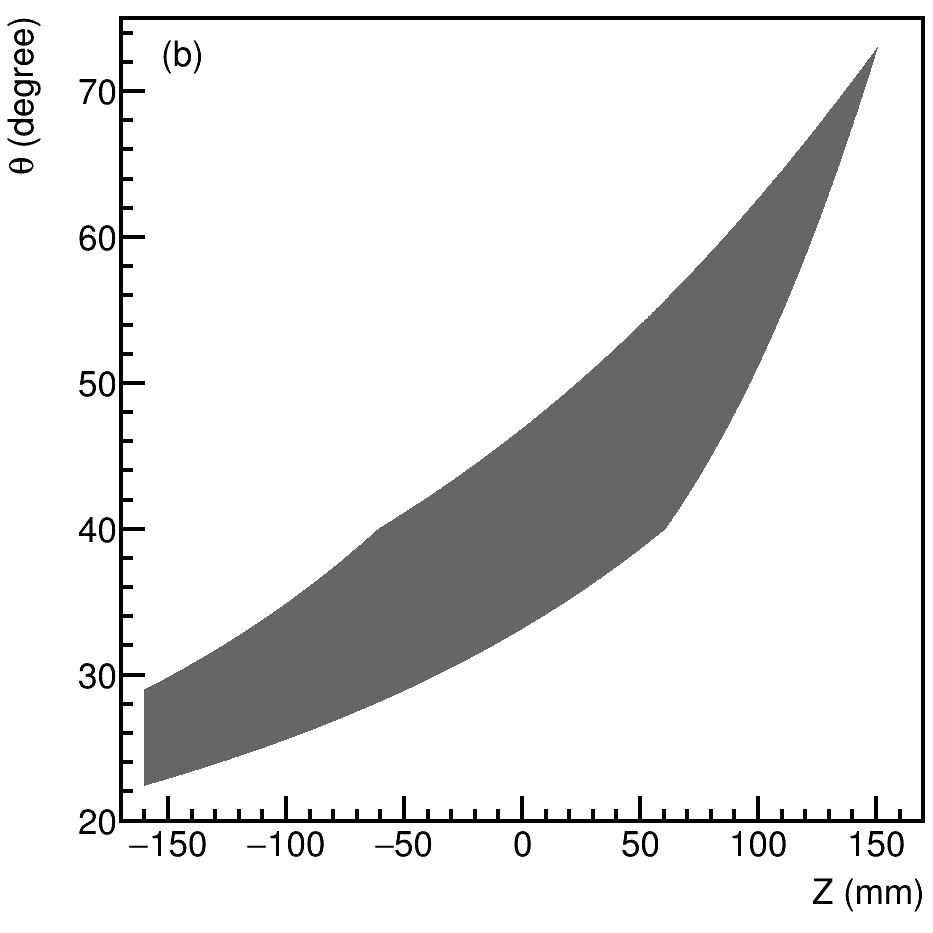}
    {acceptance}
    {Tracker acceptance as a function of fragment angle from the $Z$ axis for (a) \qty{16}{\cm} phantom, and (b) \qty{32}{\cm} phantom. Acceptance was computed for particles originating from a point directly on the $Z$ axis (i.e. $x=0$, $y=0$).}{Tracker acceptance as a function of fragment angle.}
    
    Parameters of the radiation beam were controlled by the HIT beam delivery system. For each energy change, a new spill was conducted by the synchrotron; no reacceleration or online energy change was performed.
    
    \subsection{Data Collection}
    \label{sec:ivi_collection}
    
    Irradiations were completed in sequence from greatest to least BP depth, beginning with the \qty{16}{\cm} phantom, and followed by the \qty{32}{\cm} phantom. At each depth, six complete \qty{4.9}{\s} spills of the maximum deliverable \cion{} intensity (nominally \qty{8.00e7}{\ion \per\s}) were delivered, each comprising approximately \num{3.7e8} primary ions. As no communication between the beam delivery system and the data acquisition system was implemented, each irradiation required manual starting and stopping of the data acquisition system, which occasionally resulted in either the first or the last spill being truncated. Therefore, for each energy, either five or six spills were fully recorded, for a minimum of approximately \num{2e9} primary \cion{} ions at each BP depth. Acquired data was processed by the digital readout electronics, before being written to disk by the control and data acquisition PC for offline analysis.
    
    In the \qty{16}{\cm} phantom, a total of \num{18} BP depths were delivered, separated into three blocks: six each at approximately \qty{75}{\mm}, \qty{52}{\mm}, and \qty{27}{\mm}. Each block consisted of six adjacent energy steps from the HIT beam library, differing in BP depth by approximately \qty{0.9}{\mm} in the PMMA phantom. For the \qty{32}{\cm} phantom, a total of \num{20} BP depths, in four blocks, were delivered: six each at approximately \qty{154}{\mm} and \qty{130}{\mm}, and four each at approximately \qty{105}{\mm} and \qty{75}{\mm}. As in the \qty{16}{\cm} phantom, the blocks of six consisted of six adjacent energy steps separated by approximately \qty{0.9}{\mm}, while the blocks of four consisted of three steps separated by \qty{0.9}{\mm}, and one step separated by a further \qty{2.5}{\mm} (i.e. skipping the second- and third-lowest energy steps from a block of six). A complete listing of the beams delivered is given in \tabref{depths}.
    
    \begin{table}[!b]
        \caption[Beams delivered to the \qty{16}{\cm} phantom, and \qty{32}{\cm} phantom.]{Beams delivered to the \qty{16}{\cm} phantom (left), and \qty{32}{\cm} phantom (right). Rule lines separate the beams into blocks of similar depth.}
        \label{tab:depths}
        \begin{minipage}{.5\linewidth}
            \centering
            \begin{tabularx}{7.8cm}{C C}
                \toprule
                Energy (\unit{\MeV\per\nucleon}) & Range (\unit{\mm})\\
                \midrule
                206.91 & 78.1\\
                205.60 & 77.2\\
                204.27 & 76.4\\
                202.95 & 75.5\\
                201.62 & 74.7\\
                200.28 & 73.8\\
                \midrule
                167.66 & 54.3\\
                166.15 & 53.5\\
                164.63 & 52.6\\
                163.09 & 51.8\\
                161.55 & 50.9\\
                159.99 & 50.1\\
                \midrule
                120.45 & 30.5\\
                118.52 & 29.6\\
                116.57 & 28.7\\
                114.60 & 27.9\\
                112.60 & 27.0\\
                110.58 & 26.2\\
                \bottomrule
            \end{tabularx}
        \end{minipage}%
        \begin{minipage}{.5\linewidth}
            \centering
            \begin{tabularx}{7.8cm}{C C}
                \toprule
                Energy (\unit{\MeV\per\nucleon}) & Range (\unit{\mm})\\
                \midrule
                312.70 & 156.3\\
                311.64 & 155.4\\
                310.58 & 154.6\\
                309.52 & 153.7\\
                308.46 & 152.8\\
                307.40 & 152.0\\
                \midrule
                282.67 & 132.2\\
                281.57 & 131.4\\
                280.48 & 130.5\\
                279.38 & 129.7\\
                278.29 & 128.8\\
                277.19 & 128.0\\
                \midrule
                251.24 & 108.5\\
                250.08 & 107.7\\
                248.91 & 106.8\\
                245.39 & 104.3\\
                \midrule
                206.91 & 78.1\\
                205.60 & 77.2\\
                204.27 & 76.4\\
                200.28 & 73.8\\
                \bottomrule
            \end{tabularx}
        \end{minipage}
    \end{table}
    
    Each irradiation was completed with the smallest achievable beamspot size: between \qty{3.9}{\mm} and \qty{5.1}{\mm} FWHM for the \qty{32}{\cm} phantom, and between \qty{5.0}{\mm} and \qty{8.4}{\mm} FWHM for the \qty{16}{\cm} phantom. Larger spot size was always associated with lower primary beam energy, and therefore lower BP depth.
    
    \subsection{Data Analysis}
    \label{sec:ivi_analysis}
    
    The fundamental fIVI analysis process was the same as was reported in \textcite{hymers_intra-_2021}, combining the typical IVI closest-distance projection algorithm \autocite{kelleter_-vivo_2024, finck_study_2017, ghesquiere-dierickx_detecting_2022} with a filter placing a maximum value on the separation of the two skew lines which define the beam axis and the secondary particle path. No filter on secondary particle energy was applied, other than the minimum deposited energy required to trigger the sensor. All coincidence windows were set to \qty{31.25}{\nano\s}, corresponding to five periods of the \qty{160}{\mega\Hz} timing clock. An approximate beamspot size of \qty{4.0}{\mm} was used in the reconstruction of all data sets.
    
    Due to the presence of a baseline plateau of events remaining after the distal edge, which is expected for large phantoms where forward-focused fragments may continue a significant distance beyond the BP, a minor modification was made to the previously-described fitting process for the logistic function \eqref{eq:logistic}. Rather than extending the fit to an arbitrarily-large horizontal value beyond the phantom, the maximum horizontal value was selected as \qty{+40}{\mm}. This value was chosen as it was beyond the distal edge for all measured BP positions, but still within the field of view of the tracker, prior to beginning the decline of the distal plateau towards zero.
    
    \begin{equation}\label{eq:logistic}
        f(x) = y_0 + \frac{h}{1 + \exp{(-w (x - x_0))}}
    \end{equation}
    
    To normalize the fitting between BP depths which differed significantly in distal edge position and total collected statistics, all vertex distributions were scaled to a global maximum of \qty{250}{\vertices\per\mm} prior to fitting, rather than scaling one distribution to match the other. With this change, the identification of the local maximum for the fit range was changed to require a valley at least \num{16} (scaled) counts below the calculated maximum. \num{16} was chosen as the square root of \num{250}, rounded to the nearest integer.
    
    As the amplitude of the distal plateau varied significantly between data sets, it was not always possible to achieve good matching of both the lower asymptote $y_0$ and height $h$ of the logistic fits, following normalization. This phenomenon particularly impacted the \qty{32}{\cm} phantom, where the longer exit path of secondary particles through the phantom allowed for more deflection due to multiple scattering, leading to a larger width parameter $w$, and causing the plateau region to appear less flat. To compensate, the region compared in the range shift algorithm was adjusted to begin at the same point as the fit region, truncating the function in the $-Z$ direction, while extending \qty{100}{\mm} from the inflection point in the $+Z$ direction (beyond the end of the fit region). The logistic fit functions were translated vertically to match the lower asymptote, but the height was allowed to differ.
    
    \section{Results}
    \label{sec:ivi_results}
    
    \subsection{Vertex Distributions}
    \label{sec:ivi_dist}
    
    The distribution of reconstructed interaction vertices along the nominal beam axis ($Z$) are used in the range monitoring algorithm. These distributions are arranged into histograms with a bin size of \qty{1.0}{\mm}, regardless of the phantom size or BP depth. The histogram origin is placed at the center of the phantom, matching the lab frame coordinate system. Key features of a vertex distribution, such as the example in \figref{dist32}, include the proximal edge, the distal edge, and the distal plateau. The proximal edge appears at the left of the plot, where the interaction vertex density grows from approximately zero outside of the phantom, towards a maximum as interactions begin to occur. The distal edge appears opposite the proximal edge, where the interaction vertex density decreases. This feature is the most sensitive indicator of BP depth, and is fit using a logistic function as part of the range monitoring process. The distal plateau occurs at depths beyond the distal edge. These interaction vertices are presumed to correspond to scattered fragments, as previous simulations have shown they can be significantly suppressed using the fIVI energy filter \autocite{hymers_monte_2019}. The distal plateau then descends approximately linearly towards zero at the edge of the tracker’s field of view.
    
    \singleFig[t]{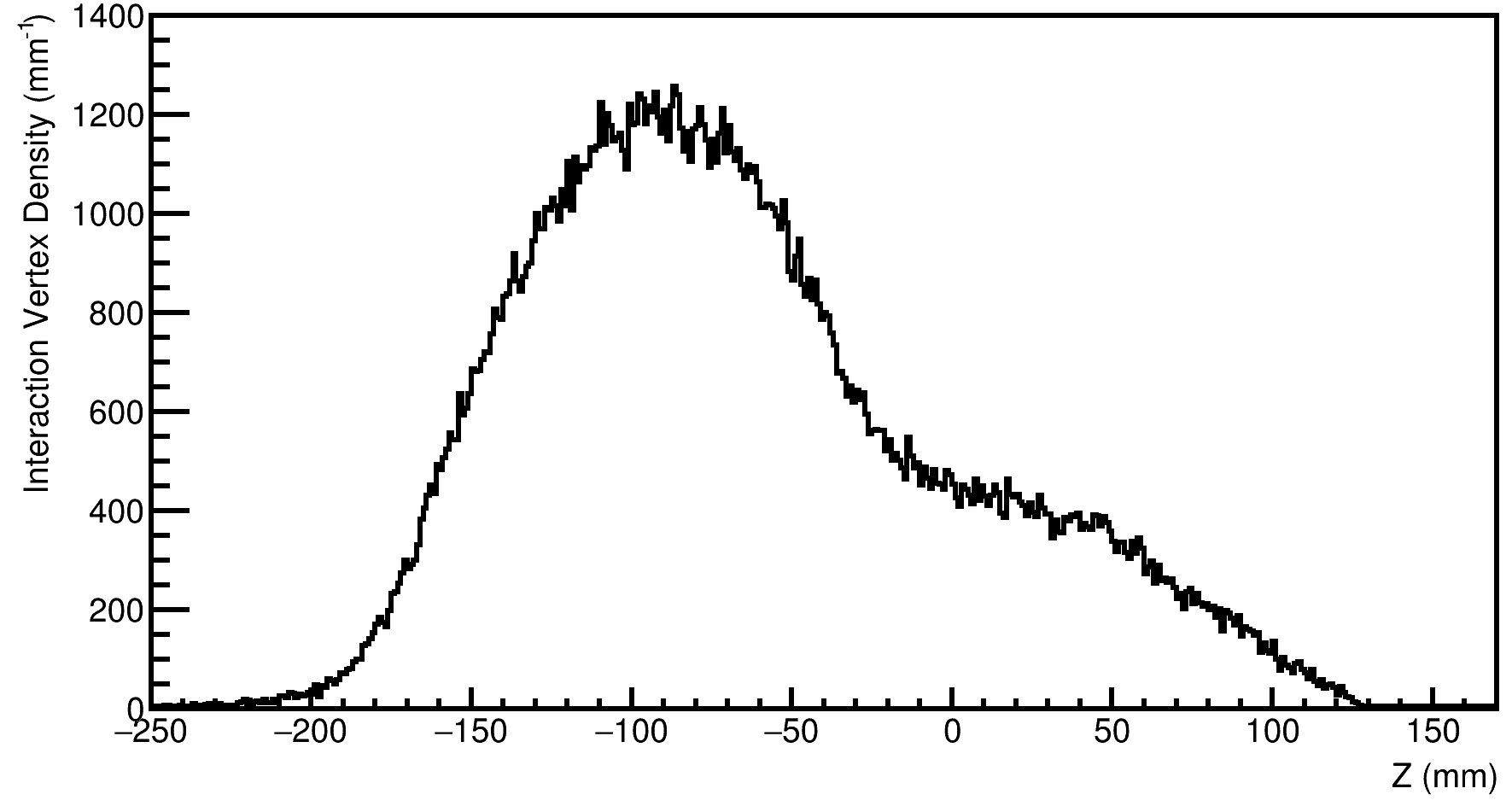}{dist32}{A typical interaction vertex distribution produced from one spill in the \qty{32}{\cm} phantom at a BP depth of \qty{156.3}{\mm}. The proximal edge is visible as a sigmoidal curve beginning at approximately \qty{-200}{\mm}, while the distal edge is centered around \qty{-50}{\mm}. The distal plateau extends until approximately \qty{+50}{\mm}, before linearly dropping to zero at around \qty{+130}{\mm}, an effect of the field of view of the fIVI tracker.}{A typical interaction vertex distribution produced from one spill.}
    
    Significant changes in the statistics collected in the vertex distribution are apparent as a function of BP depth, even for a fixed number of primary ions, as shown in \figref{bpstat}. This variation affects both the total number of reconstructed vertices and the amplitude of the local maximum. At shallower BP depths, the number of interaction vertices is decreased due to both the reduction in beam path length, which reduces the fraction of incident ions that undergo fragmentation reactions, and the increase in exit path length, which reduces the fraction of secondary fragments that exit the phantom without undergoing significant scattering. Regardless of BP depth, consistent fitting of the distal edge becomes challenging for vertex distributions with a global maximum of \qty{200}{\vertices\per\mm} or fewer.
    
    \singleFig[t]{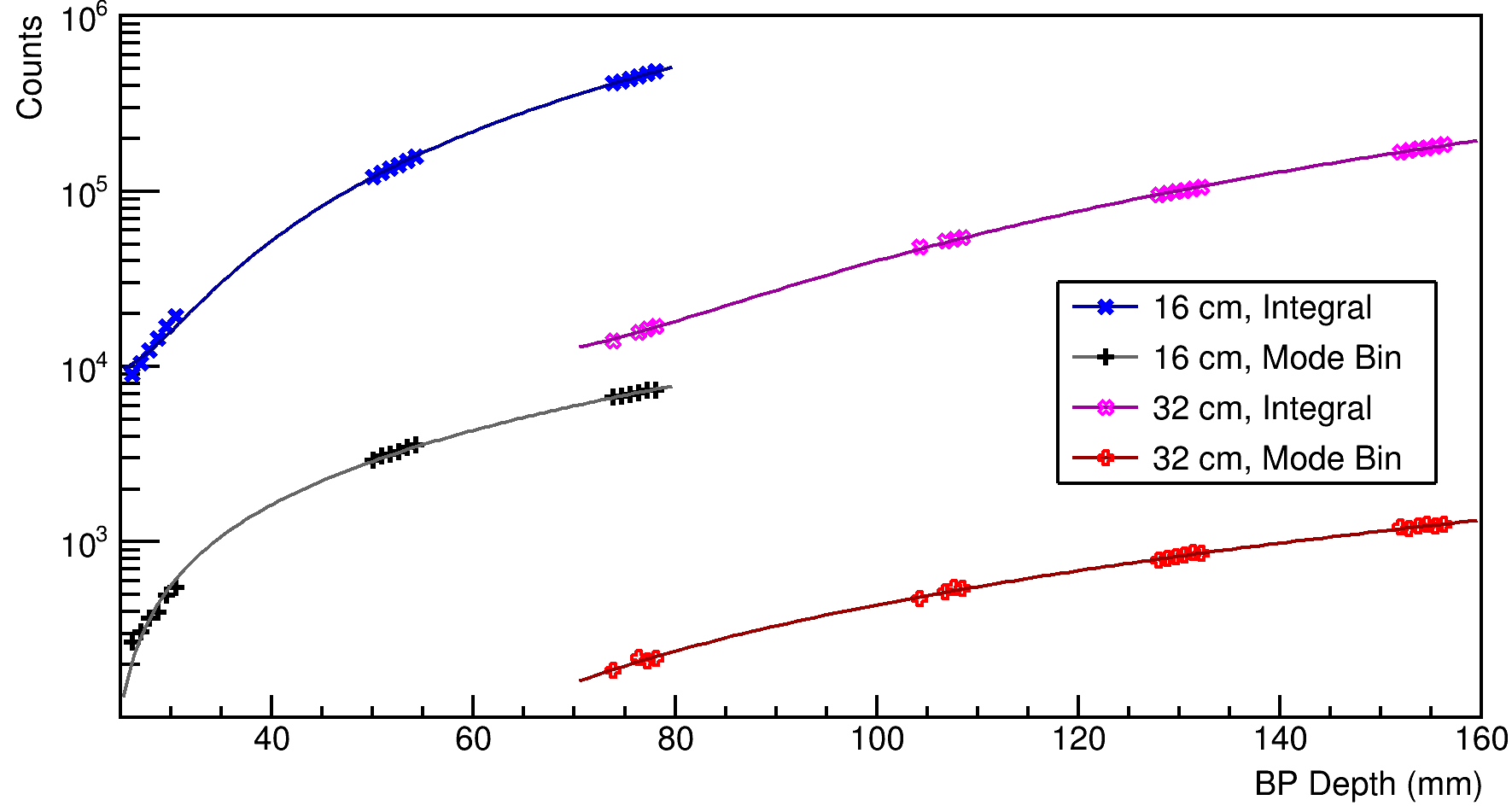}{bpstat}{Relationship between BP depth and collected statistics per spill for the complete vertex distributions (blue, magenta) and the global maximum bin used for normalization (black, red). The number of counts in the global maximum bin is an important surrogate signal for the ability to perform range monitoring: when collected statistics are so low that this amplitude drops below \qty{200}{\vertices\per\mm}, fitting the distal edge is no longer consistent. Data are presented for both the \qty{16}{\cm} phantom (blue, black), and the \qty{32}{\cm} phantom (magenta, red). Each data set is fit using a quadratic function.}{Relationship between BP depth and collected statistics per spill.}
    
    \subsection{Range Monitoring}
    \label{sec:ivi_rm}
    
    The overall goal of fIVI is to determine the difference in BP position between any two measurements, using the measured vertex depth distributions. A representative example comparing two similar BP depths in the \qty{16}{\cm} phantom is shown in \figref{cmp16}. Although the fit functions are shown in the figure along the distal edge region only, for range shift determination the lower asymptotes are extended along the positive axis. The reconstructed range difference between the two depths is the horizontal translation (evaluated in \qty{100}{\micro\m} increments) which minimizes the $\chi^2$ statistic between the two fit functions in the distal edge region, once translated vertically to match the lower asymptotes.
    
    \singleFig[t]{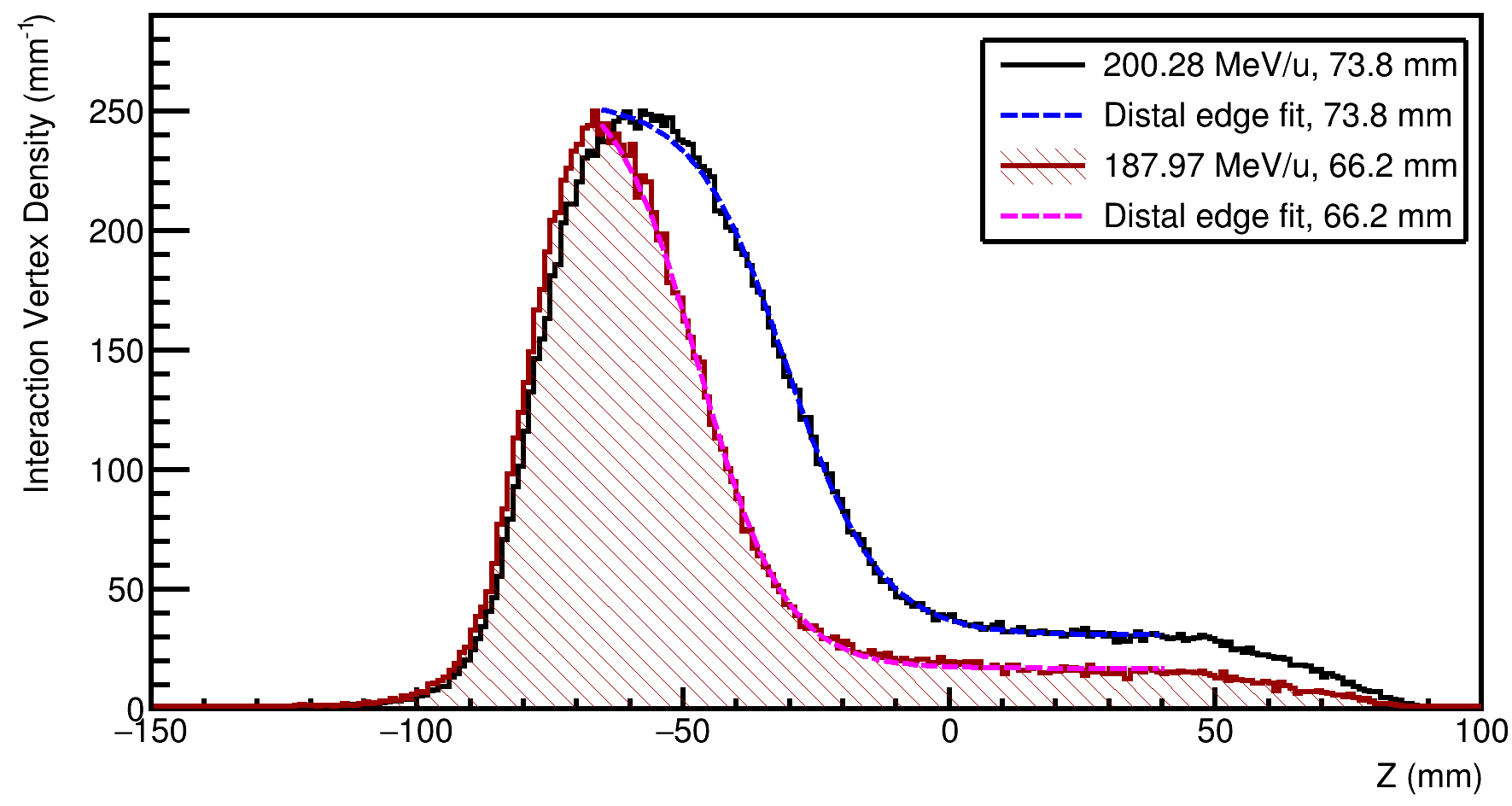}{cmp16}{Reconstructed vertex distribution comparison for full spills at BP depths of \qty{73.8}{\mm} (black) and \qty{66.2}{\mm} (red) in the \qty{16}{\cm} phantom, normalized to a global maximum of \qty{250}{\vertices\per\mm} as in the range shift determination algorithm. Fits to the distal edges are shown for the computed fit regions, in blue and magenta respectively. The proximal edges exhibit a small but detectable difference due to the slightly lower fragmentation cross section at the higher initial beam energy necessary to achieve the deeper \qty{73.8}{\mm} BP depth. Therefore, the proximal edge rises more slowly when the distal edge is deeper. The distal edges are offset due to the difference in BP position; the horizontal translation which best aligns these edges, once the lower asymptotes are aligned, yields the fIVI depth difference between the two irradiations.}{Reconstructed vertex distribution comparison.}
    
    The aggregate results of all such comparisons between complete spills for the \qty{16}{\cm} phantom are shown in \figref{summary16}. Comparisons are not performed using the datasets between \qty{26}{\mm} and \qty{31}{\mm}, due to the significantly larger beamspot sizes at very low BP depth affecting reconstruction. A comparison between two measurements is performed in two ways, with each measurement acting as both the reference measurement (i.e. the previous irradiation) and the online measurement (i.e. the current irradiation) at different times. As only the fit function of the reference measurement is used to determine the evaluation range for the  $\chi^2$ statistic, while the online measurement is translated to match, these two comparisons are not exactly equivalent. However, the magnitude of the range differences produced by permutations of the same data sets are quite comparable, with an average difference of \qty{0.1}{\mm}.
    
    \singleFig[p]{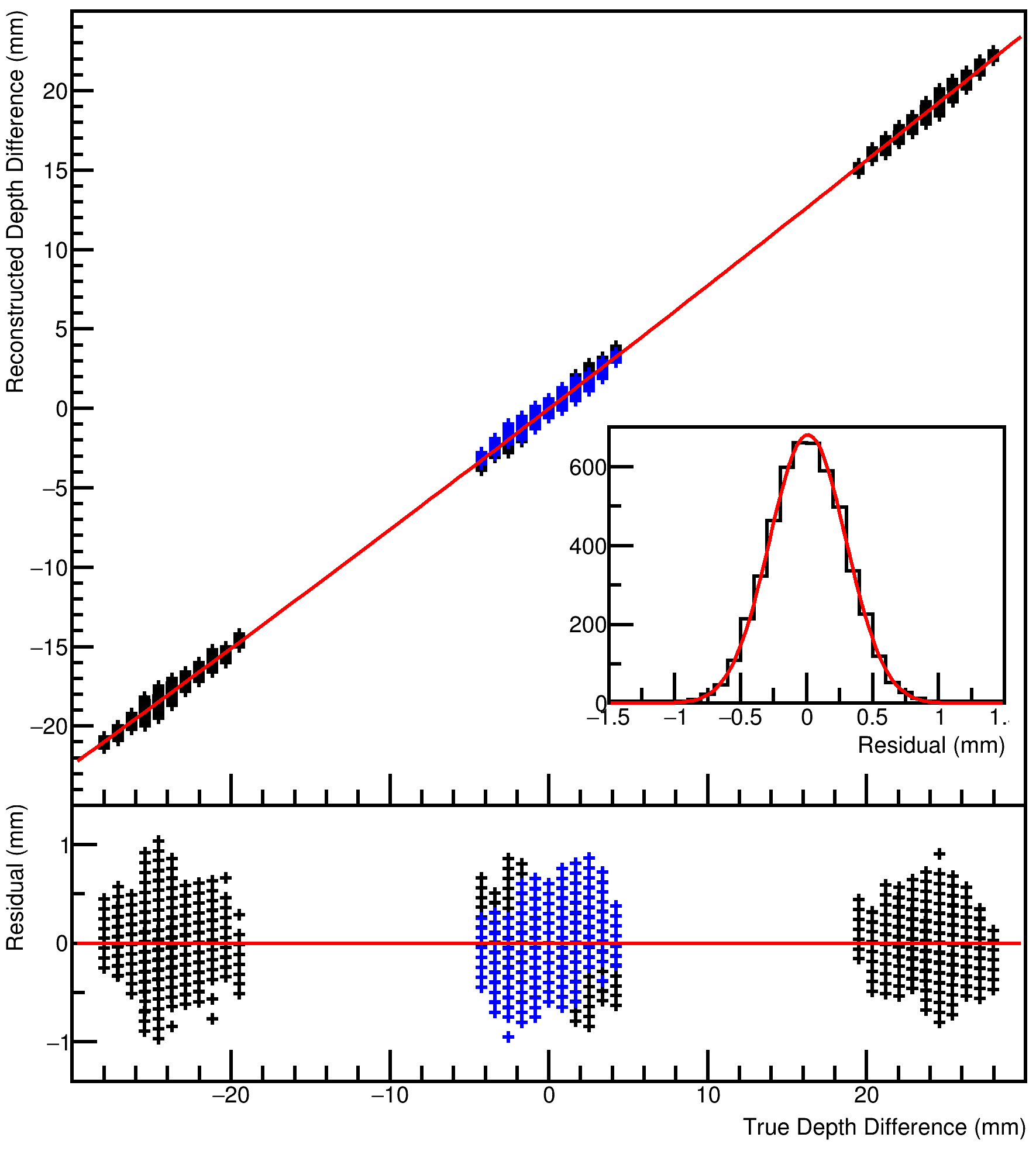}{summary16}{Aggregate result of comparisons between full spills (upper panel) in the \qty{16}{\cm} phantom. Where two block combinations overlap, each combination is plotted in a different colour. The data is fit using a second-order polynomial, with quadratic term \qty{0.000686 \pm 0.000014}{\per\mm}, slope \qty{0.76826 \pm 0.00024}{}, and intercept \qty{0.003 \pm 0.006}{\mm}. The deviation from this fit is plotted as the residual (lower panel). The distribution of the residuals about the fit (inset) forms a normal distribution, with a standard deviation of \qty{0.291 \pm 0.003}{\mm}, and worst-case deviation of \qty{1.1}{\mm}, representing the precision of this data set.}{Aggregate result of comparisons between full spills in the \qty{16}{\cm} phantom.}
    
    The overall trend in the comparisons is well-described by a linear fit, with slope \qty{0.76826 \pm 0.00029}{} and intercept \qty{0.203 \pm 0.005}{\mm}. However, when applying the linear fit, the residuals are notably asymmetric, and introducing a small quadratic term is sufficient to increase symmetry and reduce the inaccuracy in the intercept without significantly changing the slope. The introduced quadratic term is small enough that it may be entirely neglected for range differences of order \unit{\mm}, without significantly affecting the range shift determination through fIVI.
    
    The residuals about the quadratic fit, also appearing in \figref{summary16}, are symmetrically distributed about the mean, with a precision approaching the approximately \qty{200}{\micro\m} uncertainty imposed by longitudinal straggling of the treatment beam. This precision also clearly achieves the goal of sub-millimeter RM performance, exceeding the \qty{1.0}{\mm} depth in water precision with which the beam energy at HIT can be adjusted.
    
    \singleFig{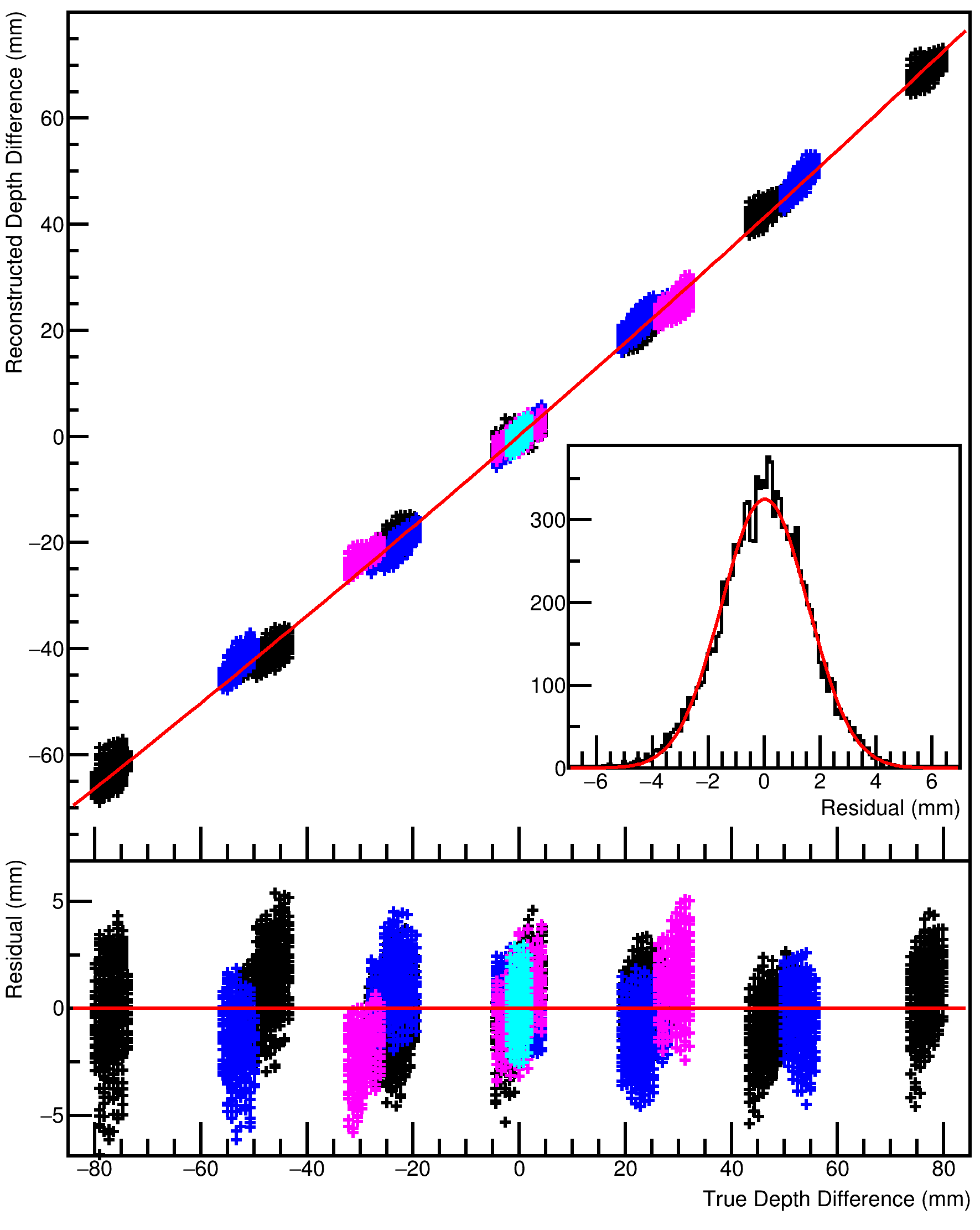}{summary32}{Aggregate result of comparisons between full spills (upper panel) in the \qty{32}{\cm} phantom. Where two block combinations overlap, each combination is plotted in a different colour. The shallowest Bragg peak, at a depth of \qty{73.8}{\mm}, is excluded, due to having a maximum below \qty{200}{\vertices\per\mm}. The data is fit using a second-order polynomial, with quadratic term \qty{0.000464 \pm 0.000008}{\per\mm}, slope \qty{0.8683 \pm 0.0004}{}, and intercept \qty{0.225 \pm 0.018}{\mm}. The deviation from this fit is plotted as the residual (lower panel). The distribution of the residuals about the fit (inset) forms a normal distribution, with a standard deviation of \qty{1.555 \pm 0.011}{\mm}, and worst-case deviation of \qty{7.0}{\mm}, representing the precision of this data set.}{Aggregate result of comparisons between full spills in the \qty{32}{\cm} phantom.}
    
    Similarly, the aggregate performance of all complete spill comparisons for the \qty{32}{\cm} phantom are shown in \figref{summary32}. Again, the range differences produced by combinations of the same two datasets result in similar performance, with an average difference of only \qty{0.1}{\mm}. The larger range of translations tested highlights the importance of the quadratic term for large deviations in BP position, and, as with the data from the \qty{16}{\cm} phantom, introducing the quadratic term results in a significant improvement in the accuracy of the intercept, as well as in the precision as defined by the residual distribution. However, the lower collected statistics of the larger \qty{32}{\cm} phantom have a negative impact on precision, causing the residual distribution to be approximately five times broader than at the \qty{16}{\cm} distribution, for full spill comparisons.
    
    Of particular note is that comparisons between each pair of data blocks, as defined in \tabref{depths}, and as shown in different colours in \figref{summary32}, may show trends which differ from the aggregate performance of the phantom. In particular, the slope and intercept of a fit function to a single data block combination may differ significantly, although the general trend of a positive slope less than unity is consistent. These differences are attributed to the correspondence between the position of the distal edge and the path length in the phantom associated with secondary fragments as they travel toward the tracker. No clear trend is observed between the BP depth and the slope, although for a given combination of blocks, an increase in collected statistics tends to bring the slope closer to unity.
    
    \subsection{Sensitivity}
    \label{sec:ivi_sensitivity}
    
    As has been previously reported, the range monitoring performance varies as a function of the collected statistics \autocite{finck_study_2017, hymers_intra-_2021}, and therefore varies as a function of both BP depth and the number of primary ions, as shown in \figref{sens}. Due to the greater collected statistics at greater BP depth, it is possible to examine aggregate precision with a number of lower primary ions by excluding irradiations at low BP depth, and focusing only on irradiations at high BP depth.
    
    \singleFig[!b]{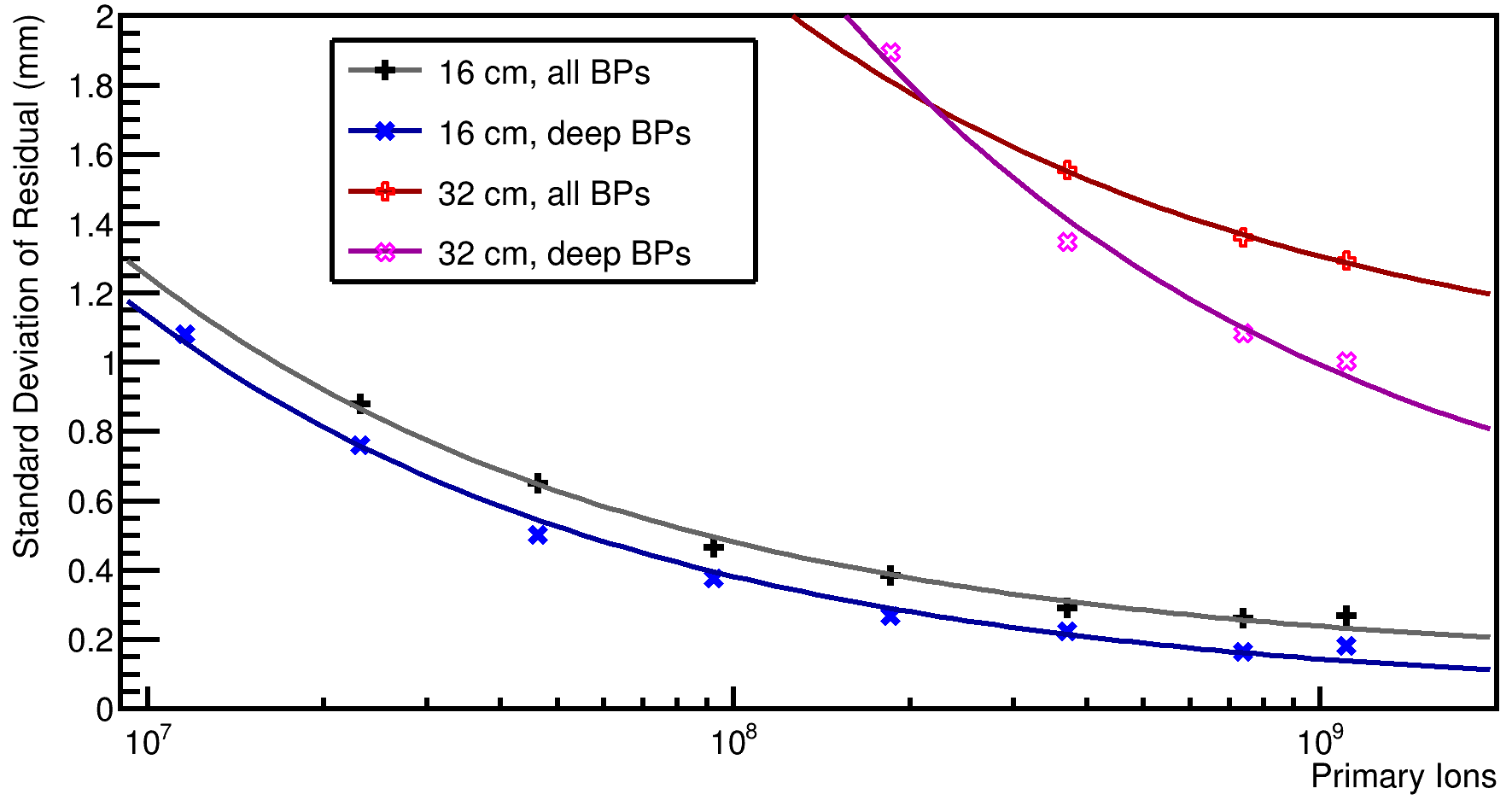}{sens}{Correlation between primary ion count and standard deviation of residual (i.e. range monitoring precision) for the \qty{16}{\cm} phantom at all depths (black) and high depths only (blue), as well as the \qty{32}{\cm} phantom at all depths (red) and high depths only (magenta). Each data set is fit with a function of the form ${f(x) = \sigma_0 + k / \sqrt{N_{ion}}}$ as modelled by \textcite{henriquet_interaction_2012} and \textcite{toppi_monitoring_2021}.}{Correlation between primary ion count and standard deviation of residual.}
    
    For the \qty{16}{\cm} phantom, optimal RM performance is already achieved when measuring a single spill, with only minor variation in the standard deviation of residual precision when combining multiple spills. The entire data set is able to be monitored down to a primary ion count of \num{2.31e7}. When focusing on only the deepest BP depths, which produce the highest statistics, the lowest primary ion count which can be monitored is a factor of two lower, \num{1.16e7}, and the precision achievable with a full spill is improved to approximately \qty{220}{\micro\m}. The consistent behaviour of the fit functions, differing by only a horizontal shift, indicates that in the \qty{16}{\cm} phantom, the reconstruction performance is limited primarily by the collectable statistics, as shown in \figref{bpstat}.
    
    For the \qty{32}{\cm} phantom, the collected statistics are insufficient to achieve optimal RM performance, even with a full spill. In fact, a full spill is the lowest number of primary ions for which RM can be reliably performed across the entire data set. Performance does improve as expected with an increasing number of primary particles, although even expanding to several spills does not approach the asymptotic precision limit. When analysis is again restricted to the deepest BP positions, similar performance improvements to the \qty{16}{\cm} phantom are observed: for a fixed number of primary ions, precision is improved, and the minimum number of primary ions which is consistently reconstructible is reduced by a factor of two.
    
    \section{Discussion}
    \label{sec:ivi_discussion}
    
    \subsection{Range Monitoring Performance}
    \label{sec:ivi_rm_perf}
    
    The prototype fIVI Range Monitoring System is capable of sub-millimetric to millimetric precision under clinical conditions, providing a highly precise method of relative range monitoring. The best performance is observed for deep BPs, for which the secondary fragment exit path length is minimized: even when controlling for the total number or peak density of secondary fragments monitored, the standard deviation of range differences is greater at shallower BP depth. The significantly different trends in precision between phantoms, directly compared for the first time in this work, highlight the sensitivity of RM performance to the exit path length. These results agree with simulation by \textcite{muraro_monitoring_2016}, which predicted nonlinear variation in most parameters of a combined fit to the proximal and distal edges of an IVI vertex distribution as a function of the exit path length of secondary fragments.
    
    RM performance is also found to vary nonlinearly with the collected statistics, with precision increasing up to vertex distribution peaks of \qty{7000}{\vertices\per\mm}, and the minimum usable statistics between \qty{200}{\vertices\per\mm} and \qty{300}{\vertices\per\mm}. The quadratic behaviour in the collectible statistics as a function of BP depth is correlated to the linear relationship between BP depth and exit path length, and is in agreement with previous calculations \autocite{muraro_monitoring_2016}. The variation in precision with vertex density follows the same form measured by \textcite{toppi_monitoring_2021} for precision in the lateral beam position. This result also follows the same form found in simulations by \textcite{henriquet_interaction_2012}, although the maximum achievable precision and point of approach to the asymptote differ significantly. These differences are attributable to overestimation of the secondary fragment production in simulations, as well as variation in the size and positioning of sensors, which suggests that improvement in precision may be possible with optimized sensor positioning.
    
    The overall trend of the relationship between the true range difference and the range difference as determined by fIVI has been found to be nonlinear with a small quadratic term. This result matches that of \textcite{finck_study_2017}, who investigated the distal edge position as a function of BP depth. However, the quadratic term observed in this work is sufficiently small that the loss in accuracy from employing a linear model is negligible. For instance, for depth differences of less than \qty{1.0}{\cm}, neglecting the quadratic term changes the calculated depth difference by less than \qty{100}{\micro\m}. Therefore, when monitoring such small depth differences, such as intra-fraction RM for adjacent isodepth slices, or inter-fraction RM for the same raster point, acceptable performance may be achieved using a purely linear model.
    
    Furthermore, in the case of comparisons between the same or adjacent raster points, expected to differ by \qty{1.0}{\mm} or less, the systematic error introduced by assuming a slope of 1 is also sufficiently small to be neglected. While this error is larger than that introduced by neglecting the quadratic term, on the order of \qty{100}{\micro\m\per\mm}, it is still significantly smaller than the millimetric precision with which BP depth can be varied at a clinical facility such as HIT. Therefore, if the detection of deviations is limited to range differences large enough to be corrected by a change in primary beam energy, the adoption of a linear model with slope 1 is a permissible simplifying assumption for RM of similarly-located BPs.
    
    As the \qty{16}{\cm} phantom is of similar dimensions to those used in previous studies, a direct comparison of the RM performance is possible. Previous studies have demonstrated the ability to reliably distinguish BP depth differences on the order of millimeters under clinical conditions \autocite{finck_study_2017, gwosch_non-invasive_2013}. This work achieves the same precision, and demonstrates an extension to shallower BP depths, with millimetric precision for BP depth as low as \qty{50}{\mm}, less than the \qty{80}{\mm} phantom radius, while prior studies have only demonstrated this precision at depths of \qty{100}{\mm} or beyond. These shallow BP positions are more relevant to clinical treatment plans, in which tumours are often approached along the shortest path length in order to maximize sparing of healthy tissue. This extension of prior performance reinforces the suitability of fIVI for a variety of clinical applications.
    
    \subsection{Clinical Applicability}
    \label{sec:ivi_clin}
    
    One major clinical goal of fIVI is to allow the use of RM for every fraction in a treatment plan, allowing the shrinking of margins through more precise monitoring of the BP position \autocite{traini_review_2019}. With this work, experimental evidence exists that IVI is capable of millimeter to sub-millimeter precision in monitoring a variety of BP depths, in cases such as pediatric patients or the adult head and neck. This precision is limited to the most intense spots, with primary ion counts on the order of \num{e6} to \num{e7}. However, this performance is already sufficient to perform inter-fraction monitoring, and apply the benefit of fIVI to all fractions, if even a single spot in the treatment plan achieves this intensity. 
    
    This work provides strong evidence that the relationship between the IVI-calculated range difference and the true range difference depends on the specific geometry of the patient or phantom and the tracker, through the direct comparison of two different phantom geometries. As the goal of fIVI is not only to detect a range error, but also to provide an estimate of its magnitude, measurements of the linear and quadratic coefficients for each patient and tumour are therefore required for maximum accuracy. One possible method of evaluating these coefficients is to use data from the first treatment fraction, for which absolute range verification could be confirmed through concurrent use of another RM technique, such as PET, as has been suggested in the literature \autocite{fischetti_inter-fractional_2020}. Determination of the patient-specific correction factor would then allow high-accuracy determination of range errors, which would be valid for a wide range of possible error magnitudes.
    
    While maximum performance requires determination of the patient- and tumour-specific linear and quadratic coefficients for each treatment plan, it is possible that sufficient accuracy for routine clinical use may be achieved in a patient-independent manner. As the treatment beam energy, and therefore depth, can only be adjusted with limited precision (typically millimetric), sufficient accuracy for small range shifts may be achieved using the patient-independent approximation of a linear model with slope \num{1} and intercept \num{0}. In the \qty{16}{\cm} phantom, this approximation introduces an inaccuracy of less than \qty{0.5}{\mm} for range errors of less than \qty{2}{\mm}, while in the \qty{32}{\cm} phantom, this same accuracy limit is maintained for range errors of less than \qty{3}{\mm}. Notably, range errors corresponding to small-magnitude range shifts, specifically those addressed by current treatment planning safety margins, are expected to be the most frequent finding of fIVI \autocite{kelleter_-vivo_2024}. Further studies are required to confirm whether this approximation remains valid in all cases.
    
    For clinical inter-fraction RM, it is also necessary to account for uncertainties posed by small changes between fractions in alignment of the patient and monitoring system. In particular, differences in longitudinal patient positioning may produce a shift in the absolute distal edge position, merely because of a millimetric difference in the separation between the patient and the beam nozzle. While the energy loss in an extra few millimeters of air would not meaningfully impact the outcome of treatment, a distal-edge-only approach to IVI would detect a range error. To combat this issue, alignment corrections could be made using the proximal edge of the vertex distribution, which has been reliably demonstrated here, and previously, to behave similarly for BP depths differing by a few millimeters \autocite{finck_study_2017, gwosch_non-invasive_2013, hymers_intra-_2021}. Such an improvement could perhaps use a combined logistic fit to the proximal and distal edges, as proposed by \textcite{piersanti_measurement_2014}; however, further investigation is needed to determine whether such a fit may be affected by inhomogeneities superficial to the observable distal edge region, and the effect of changes in acceptance angle as a consequence of alignment errors on the accuracy of a proximal edge alignment. As the proximal edge shape observed in this work is qualitatively similar for range differences of less than \qty{4.5}{\mm}, such an alignment is expected to be feasible for inter-fraction comparisons of the same raster point.
    
    While this work has achieved millimetric precision with as few as \num{e7} primary \cbase{} ions for centrally-located Bragg peaks in phantoms with similar dimensions to the head or neck, increasing the collectable statistics further would have a direct impact on both the precision with which RM could be performed for these most intense spots, and the number of spots capable of achieving millimetric precision. Optimizations such as reducing the distance between sensor layers, increasing sensor sizes, or using a smaller enclosure -- facilitating tracking at a smaller off-axis angle -- are expected to allow for sufficient improvement to allow clinical monitoring of these high-statistics raster points when using a single two-layer tracker. The sensors used in the fIVI Range Monitoring System have also been designed for the construction of dense arrays, creating large, nearly-continuous sensitive areas composed of hundreds of sensors \autocite{heuser_technical_2013}. Therefore, there is no obstacle in scaling the number of sensors to meet any clinically-relevant use case.
    
    In addition to inter-fraction monitoring, it is also desirable to achieve intra-fraction monitoring, which would allow the detection of undesirable patient motion which occurs during irradiation. To achieve this goal, monitoring must be sensitive to many individual points in a raster scanned treatment. Typical scan points receive \numrange{e4}{e6} primary ions per fraction, with higher values often being located more distally on the tumour \autocite{kramer_treatment_2000, finck_study_2017}. The orders of magnitude increase in the vertex reconstruction rate required to achieve this performance likely exceeds that which can be achieved through optimization of the single tracker in the prototype system. Consequently, moving to larger, specialized sensor arrays may be unavoidable \autocite{finck_study_2017, henriquet_interaction_2012}. In this case, the use of large-area sensors poses significant benefits over the smaller sensors used in most previous IVI studies. With large-area sensors like those used in this work, the total sensitive area may be made very large while maintaining a modest number of total sensors, which meaningfully reduces the complexity of control, synchronization, and management of the RM system.
    
    Scaling to larger sensitive areas is also expected to be required to apply fIVI for abdominal tumours, due to the reduced RM precision observed in the larger phantom. The reduced performance is related to the large exit path length of secondary fragments, which both reduces the number of detectable fragments and the precision associated with each fragment, due to the increased influence of multiple Coulomb scattering \autocite{ghesquiere-dierickx_investigation_2021}. In addition to the aforementioned increases in sensitive area, another option is to explore aggregate monitoring, combining the results from multiple raster points at the same BP depth to increase the collectable statistics, without requiring additional sensors. Further analysis is required to determine the impact this method will have on the precision of range monitoring. Data have already been collected for isodepth surfaces under the same conditions as the presented results, and analysis is ongoing.
    
    Future work based on the prototype fIVI Range Monitoring System will focus on optimizing the tracker and enclosure design to improve the collectable statistics with the existing sensors. For the most challenging irradiations, additional sensors or trackers could be added to the setup, to further improve RM performance. Further tests are also planned to explore more clinically-relevant anthropomorphic phantoms, and the impact of the associated inhomogeneities on range monitoring performance. 
    
    While all of the analyses discussed in this work were completed offline, after all irradiations had been delivered, there is in principle no obstacle to employing online monitoring, comparing a previous raster spot or fraction to the current spot while delivery is ongoing. Establishing online range monitoring would enable fIVI to play an active role in the quality assurance of beam delivery, allowing treatment to be paused or aborted if range errors are detected. Such an outcome would significantly reduce the potential harm to patients, and in particular radiosensitive organs at risk, through early detection of errors in BP positioning. Furthermore, with the comparison between BP positions made by fIVI, monitoring provides feedback on not only the presence of errors in BP position, but also the approximate magnitude of the error. This additional sensitivity may help guide clinicians in investigating the source of a range error, or may even facilitate the adaptation of the treatment plan based on the reported range error. Such an adaptation could ultimately be carried out in the treatment room, allowing the scheduled treatment to be completed rather than aborted when a range error is detected.
    
    \section{Conclusion}
    \label{sec:ivi_conc}
    
    The prototype fIVI Range Monitoring System represents the successful first application of large-area silicon sensors to range monitoring in carbon ion radiotherapy. This device is capable of consistently measuring range shifts in Bragg peak position for clinically-relevant \cion{} beams in human scale PMMA phantoms. In a \qty{16}{\cm} diameter phantom, \num{4970} comparisons for two Bragg peak depths between \qty{50.1}{\mm} and \qty{78.1}{\mm} measured depth differences between full spills with a standard deviation of \qty{291 \pm 3}{\micro\m}, while a standard deviation of less than \qty{1.0}{\mm} was maintained for irradiations of \num{2.3e7} primary ions, approaching the ability to monitor individual beam spots within a scanned treatment. This result represents the first experimental extension of IVI to shallow BPs in a human-scale phantom. For the larger \qty{32}{\cm} diameter phantom, statistics limited the achievable precision to the order of \qty{1.0}{\mm} standard deviation, even at full spills, indicating that even larger sensors are needed to achieve clinically-useful monitoring of treatments in the abdominal region. Clinical implementation of fIVI is projected to allow robust monitoring, supporting the consistent delivery of each fraction in a treatment plan. Further investigation of this system is underway to explore improvements in acceptance, to facilitate improved collection of secondary charged fragments in the challenging environment posed by large or inhomogeneous phantoms and patients.
    
    % References
    \printbibliography[title=References]
    
\end{document}